\documentclass{edp-conf}
\usepackage{graphicx}

\def\spose#1{\hbox to 0pt{#1\hss}} 
\def\lta{\mathrel{\spose{\lower 3pt\hbox{$\mathchar"218$}} 
        \raise 2.0pt\hbox{$\mathchar"13C$}}}      
\def\gta{\mathrel{\spose{\lower 3pt\hbox{$\mathchar"218$}} 
        \raise 2.0pt\hbox{$\mathchar"13E$}}} 

\begin{document}

\TitreGlobal{Physics near a black--hole horizon}

\title{Physics near a black--hole horizon} 

\author{Jean-Pierre Lasota}
\address{Institut d'Astrophysique de Paris, 98bis Boulevard Arago, 75014 Paris}

\maketitle

\begin{abstract}
I discuss several general-relativistic effects that are likely to be of 
interest in the astrophysics of black holes and neutron stars
\end{abstract}

\section{The horizon}

The principal attribute of black holes, which makes them different from
any other compact object, is the absence of a `hard' surface.  Finding
the signature of this unique property in some observed systems would
be the ultimate proof of black-hole existence.  Until recently only
the mass of a compact body in excess of the upper limit for neutron
stars or, in the case of   galactic nuclei, a mass concentration in
excess of the upper limit for a sufficiently long-lived stellar
cluster, could be used to decide that such a compact object belongs to
the black-hole family. Some people prefer to call such objects
black-hole `candidates'. We know now almost fifteen close binary
systems, in which the compact object is believed to be a black hole
(see e.g. Charles 2001) and a large number of galaxies (including our
own) are believed to contain supermassive black holes in their centers
(e.g. \cite{magor}).  Are these objects true black holes? In the
recent years Ramesh  Narayan and his collaborators have established
that X-ray luminosities of quiescent low-mass X-ray binary transient
systems (also known under the name of `Soft X-ray Transients' or
`X-ray Novae')  containing compact objects selected as black holes
because of their high masses, are much lower than the corresponding
luminosities of system known to contain neutron stars and attributed
this difference to the presence of an event horizon in the
mass-selected objects (\cite{ngm}; \cite{men}; \cite{garcia}).  In
other words, black hole `candidates' would be true black holes.  After
some initial confusion about what the data are showing
(see e.g. \cite{chw}) a suggestion by Lasota \& Hameury (1998) that  a
luminosity vs period diagram is the most sensible way of deciding
about differences between various systems has been adopted. Then, it
appeared that although the luminosity difference between neutron stars
and black-hole candidates is flagrant, neutron stars  are nevertheless
fainter than predicted by the model (\cite{men}). The model in
question assumes that the inner regions of the accretion flow in
quiescent X-ray transients form an ADAF (see \cite{l99} for a
review). In such a radiatively inefficient flow most of the energy
would be forever swallowed  by a black hole but re-radiated from the
surface if the accreting compact body is a neutron star, hence the
difference in luminosities for the same accretion rate. It seems,
however, that in neutron-star quiescent transient  systems most of
the accretion energy {\sl is not} emitted from the surface of neutron
stars (i.e. the accretion efficiency is much lower than the
`standard' value $\approx 0.1$; see \cite{men} for details).

Abramowicz \& Igumenshchev (2001) suggested that the observed
differences between quiescent luminosities of accreting black holes
and neutron stars can be explained by the presence of a CDAF
(Convection Dominated Accretion Flow; see Narayan, Igumenshchev  \&
Abramowicz 2000) in such systems. They found that for low  viscosities
accretion flows around compact bodies form ADAFs  only in their
innermost regions but are convectively dominated  at radii $R\gta
10^2R_{\rm S}$ (where $R_{\rm S}=2GM/c^2$ is  the Schwarzschild
radius). In such flows emission comes mostly  from the convective
region; the radiative efficiency is  independent of accretion rate and
equals $\varepsilon_{\rm BH}=  10^{-3}$. Assuming that the efficiency
of accretion onto a neutron star is $\varepsilon_{\rm NS}\approx 0.1$
one obtains the observed ratio between black-hole and neutron-star
luminosities.  Unfortunately this cannot be the correct explanation of
the luminosity difference because, as mentioned above, neutron stars
in quiescent transient  systems do not seem to accrete with a 0.1
efficiency (\cite{men}).

Several suggestions have been put forward to explain this low
efficiency. Winds from ADAFs, suggested by Blandford \& Begelman
(1999; Paczy\'nski 1998 and Abramowicz, Lasota  \& Igumenshchev 2000
question the validity of the arguments presented in this article but
their arguments do not preclude the existence of winds of e.g. magnetic
origin) and modeled by Quataert \& Narayan (1999) are not sufficient
to explain neutron-star's low accretion efficiency. Menou et
al. (1999) proposed that the action of a magnetic propeller could be
the answer, but a compelling signature of this effect has yet to be
found. Finally, a simple and drastic suggestion was put forward by
Brown, Bildsten, \& Rutledge (1998): most (or all)  of the quiescent
X-ray luminosity is not due to accretion but results from cooling
of the neutron-star crust heated by nuclear reactions.

The crust-cooling model, however, is apparently contradicted by the
observed  variations of the quiescent luminosity on a time-scale of
years  (Rutledge 2001a,b). Luminosity variations are observed also in
quiescent black-hole systems (see e.g. G2001) which  would suggest a
common origin. Attempts to ascribe quiescent X-ray  luminosity in
these latter systems to black-hole's companions  (Bildsten \& Rutledge
2000) are theoretically unsound (Lasota 2000)  and have been refuted
by observations (G2001). Lasota (2000)  found that the correlation
between quiescent luminosities and orbital periods of black-hole
transients  (three at that time) can be explained by a simple
disc+ADAF model.  However, {\sl Chandra} observations (G2001) showed
that  things are more complicated, mainly because of the luminosity
variations  mentioned above.

Therefore, although observations seem to imply the presence
of event horizons in bodies with masses higher than the 
neutron-star maximum mass, the uncertainties which
still haunt accretion physics do not allow us to draw any definite 
conclusions in the matter.

\section{The light}

Light trajectories are strongly deflected close to the black hole
surface.  The best and the most beautiful images of a black hole
surrounded by a luminous Keplerian disc were produced by
\cite{jam}. Marck (1996; see also Hameury, Marck \& Pelat  1994)
showed that using the Kerr - Schild coordinate  system greatly
simplifies the form of the geodesic equations in the Kerr metric and
applied this form in numerical computations of black hole images.
This method was also used to calculate spectra emitted near rotating
black holes (\cite{hmp}). 
Jean-Alain Marck's work on images of a thin disc around a Kerr 
black hole was cut short by his untimely death in May 2000. 

In the near future it might be possible to see images
of black holes observed in X-rays by the {\sl MicroArcsecond 
X-ray Imaging Mission} (MAXIM; http://maxim.
gsfc.nasa.gov).

\section{The disc}

The best known general-relativistic effect in accretion disc structure
is the existence of an Innermost Stable Circular Orbit (ISCO). It
is the orbit where the Keplerian angular momentum has a minimum  and
marks the inner edge of a {\sl Keplerian} accretion disc. It has  been
shown that it is also the place where, for geometrically thins discs,
`viscous' stresses approximately  vanish ensuring flat angular
momentum profile down all over to the surface  of the black hole
(e.g. Abramowicz \& Kato 1989). The accretion
efficiency  is thus determined by the binding energy at this
orbit. This conclusion has been recently challenged (see Hawley \&
Krolik 2000 and references therein) but, as recently recalled by
Paczy\'nski (2000), it is the angular momentum conservation that
requires a `no-torque inner boundary' for geometrically thin accretion
discs around black-holes. Indeed, angular momentum conservation
implies that at the sonic ring (which almost coincides with the inner
disc's boundary):
\begin{equation}
{ v_r \over v_s} = 1 \approx \alpha ~ { H_{in} \over r_{\rm in} } ~ 
{ l_{\rm in} \over l_{\rm in} - l_0 } ;
\hskip 1.0cm r = r_{\rm in},
\label{son}
\end{equation}
where $ v_r $ is the radial velocity, $ v_s$ is the sound
velocity, $ l(r) $ is the specific angular momentum
at radius $ r $ ($l_{in}$ is the specific angular momentum at
the inner disc's edge $r_{\rm in}$, and $ l_0 $ is an integration 
constant equal
to the angular momentum at the inner flow boundary, i.e. at
the black-hole surface).

In a thin disc $ H_{in} / r_{in} \ll 1 $, Eq. (\ref{son}) implies
that for small viscosities, i.e.  for $ \alpha \ll 1 $, 
$ (l_{in} - l_0 ) / l_{in} \ll 1 $, i.e. the specific angular
momentum at the sonic ring is almost equal to its value at the
horizon.

In a stationary disc (accretion rate $\dot M= const.$) the torque $ g $ 
has to satisfy the equation of angular momentum conservation:
\begin{equation}
g = \dot M \left( l - l_0 \right) , \hskip 1.0cm 
g_{in} = \dot M \left( l_{in} - l_0 \right) 
\label{g}
\end{equation}
which shows that the `no-torque inner boundary condition' is an 
excellent approximation for a thin, low viscosity disc.  
However, if the flow is thick, i.e.  $ H/r \sim 1 $, and 
viscosity high ($ \alpha \lta 1 $), the angular momentum varies 
also between $r_{\rm in}$ and $r_{\rm S}$.  

These simple arguments have been confirmed by numerical calculations  
(Chen, Abramowicz \& Lasota 1997; Armitage, Reynolds \& Chiang 2001).
This does not mean that no coupling is possible between a thin disc
and a black hole -- this only means that this coupling
 has to be global (`non-viscous')
(see e.g. Blandford \& Znajek 1977; King \& Lasota 1977).

Non-keplerian `discs' can extend down to the IBCO 
(`B' stands for {\sl bound}; see Abramowicz \& Lasota 1980, 
for the effect this has on the maximum angular momentum of an accreting 
black-hole). Still closer to the black-hole one finds in the IKCO 
(where `K' stands of `Keplerian'), in other words the PCO (Photon 
Circular Orbit). The spatial 2D sphere at the locus of 
this orbit has strange properties discovered by Abramowicz \& 
Lasota (see 1997 and references therein), which will be mentioned in Sect.
\ref{ns} 

\section{QPO's, black holes and neutron stars}
\label{ns}

Timing observations of accreting neutron stars and black holes
in Low Mass X-ray Binaries (LMXBs) reveal pairs of simultaneous
high frequency ($\nu \gta 50 Hz$) Quasi-Periodic Oscillation (QPOs), 
which appear as peaks in the power spectrum (the Fourier transform
of time variations) of the observed X-ray flux (see van der Klis 2000;
Strohmayer 2001a,b). 

\begin{figure}
\centering 
{\includegraphics[scale=1,totalheight=6.5cm]{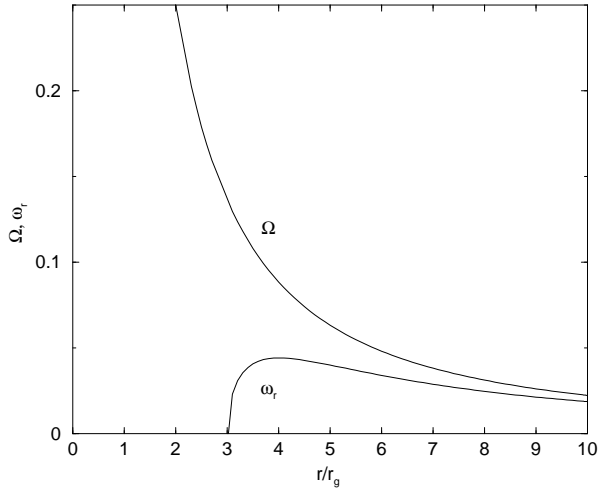}}
\caption{Plots of the
epicyclic frequency, $\omega_r$, and the orbital frequency in circular
orbits, $\Omega$, both in units of $c/r_g$, are shown as a function of
the circumferential radius in units of the gravitational radius
$r_g=2GM/c^2$ for the Schwarzschild metric (from Klu\'zniak \& 
Abramowicz 2001).}
\label{ke}
\end{figure}

Several ideas have been put forward to explain the double frequency
peak phenomenon (see references in Strohmayer 2001a). Here I will
mention only a recent suggestion by Klu\'zniak \& Abramowicz (2001),
who attribute the double peaks to a purely general-relativistic
effect. As pointed out long time ago by Kato \& Fukue (1980), the
strong deviations from the 1/r law due to the presence of a scale
($R_S$) when the gravitational field of a spherical body is described
by General Relativity, imply that the epicyclic frequency $\omega_r =
\left(r^{-3}{\rm d} l^2/{\rm d} r\right)^{1/2}$ is different  from the
Keplerian frequency $\Omega_{\rm K}=\left(GM/r^3\right)^{1/2}$ and has
a maximum (see  Fig. \ref{ke}).  For a Schwarzschild black hole this
maximum is at $r_{\rm max}= 4r_S$,  ($\omega_{max}=\Omega(4r_S)/2$; at
$r_{\rm ISCO}= 3r_S$,  $\omega_r=0$).  Therefore near the ISCO:
$\Omega_K(r)/\omega_r(r)\rightarrow\infty$, as $r\rightarrow r_{\rm
ISCO}$  which, because the (radial) epicyclic motion is anharmonic,
makes  possible prominent  1:2, 1:3 resonances between $\Omega_K(r)$
and  $\omega_r$.  Klu\'zniak \& Abramowicz (2001) suggest that the
high frequency  QPOs are caused by such resonances. Strohmayer (2001a)
observed a 450 Hz QPO  simultaneous with the previously known 300 Hz
oscillation. The  two frequencies are in a 3:2 ratio which could result
from either the 1:2 or 1:3 resonances ($\Omega=300$ Hz, $\Omega +
\omega_r$=450Hz, or $\Omega=$ 450Hz,  $\Omega - \omega_r$=300Hz) as
predicted by the model (Abramowicz \& Klu\'zniak 2001). However,
explaining the pair  of $\sim 40$ Hz and 67 Hz QPOs observed in GRS
1915+105 (Strohmayer 2001b), would probably require higher resonances.

Recently Heyl (2000) claimed that general-relativistic effects play
an important role in the evolution of QPOs observed during type 1
X-ray bursts occurring at the surface of accreting neutron stars.
In particular he found that the centrifugal force reversal
at the locus of the circular photon orbit (Abramowicz \& Prasanna (1990)
is of importance. However, as shown by Abramowicz, Klu{\'z}niak
\& Lasota (2001; see also Cumming et al. 2001) this claim is erroneous.

\section*{Acknowledgment}
I am grateful to the Physics Department of the Technion in
Haifa for hospitality during the writing of this article.

\end{document}